\pdfoutput=1

\documentclass[11pt]{article}

\usepackage[preprint]{acl}

\usepackage{times}
\usepackage{latexsym}

\usepackage[T1]{fontenc}

\usepackage[utf8]{inputenc}

\usepackage{microtype}

\usepackage{inconsolata}

\usepackage{graphicx}

\usepackage{booktabs}
\usepackage{array}
\usepackage{multirow}
\usepackage{makecell}
\usepackage{xcolor}

\title{MindChat: Enhancing BCI Spelling with Large Language Models \\ in Realistic Scenarios}

\author{
  \textbf{Jiaheng Wang\textsuperscript{1,2,3}},
  \textbf{Yucun Zhong\textsuperscript{1,2,3}},
  \textbf{Chengjie Huang\textsuperscript{1,2,3}},
  \textbf{Lin Yao\textsuperscript{1,2,3}}
\\
  \textsuperscript{1}The College of Computer Science, Zhejiang University, Hangzhou, China. \\
  \textsuperscript{2}The MOE Frontiers Science Center for Brain and Brain-Machine Integration, Zhejiang University, Hangzhou, China. \\
  \textsuperscript{3}The Nanhu Brain-Computer Interface Institute, Hangzhou, China.
\\
  \small{
    \textbf{Correspondence:} \{\href{mailto:lin.yao@zju.edu.cn}{lin.yao}, \href{mailto:Jiaheng-Wang@zju.edu.cn}{Jiaheng-Wang}\}@zju.edu.cn
  }
}

\begin{document}
\maketitle
\begin{abstract}
Brain-computer interface (BCI) spellers can render a new communication channel independent of peripheral nervous system, which are especially valuable for patients with severe motor disabilities. However, current BCI spellers often require users to type intended utterances letter-by-letter while spelling errors grow proportionally due to inaccurate electroencephalogram (EEG) decoding, largely impeding the efficiency and usability of BCIs in real-world communication. In this paper, we present MindChat, a large language model (LLM)-assisted BCI speller to enhance BCI spelling efficiency by reducing users' manual keystrokes. Building upon prompt engineering, we prompt LLMs (GPT-4o) to continuously suggest context-aware word and sentence completions/predictions during spelling. Online copy-spelling experiments encompassing four dialogue scenarios demonstrate that MindChat saves more than 62\% keystrokes and over 32\% spelling time compared with traditional BCI spellers. We envision high-speed BCI spellers enhanced by LLMs will potentially lead to truly practical applications\footnote{\url{https://github.com/Jiaheng-Wang/ZJUBCI_SSVEP}}.
\end{abstract}

\section{Introduction}
Brain-computer interfaces (BCIs) enable direct communication and control between brains and external devices \citep{Wolpaw2002}, rendering great opportunities for patients suffering from various neurological disorders to replace, restore, supplement, or improve their once-impaired bodily functions \citep{McFarland2017}. Past decades have witnessed a rapid development of both noninvasive and invasive BCIs wherein BCIs using noninvasive electroencephalogram (EEG) signals have achieved diverse promising applications including spelling \citep{Chen2015}, stroke rehabilitation \citep{Biasiucci2018}, robotic-arm control \citep{Edelman2019}, emotion regulation \citep{Huang2023}, etc. BCI spellers based on steady-state visual evoked potentials (SSVEPs) allow users to type and communicate without the participation of peripheral nerves and muscles. Neurophysiologically, occipital EEG signals can be effectively modulated by the specific combination of frequency and phase of visual flickering stimulation (e.g. black-and-white squares) \citep{ShangkaiGao2014}, which are referred to as SSVEPs. For an SSVEP-based BCI speller, characters are encoded by flickering squares with different frequency-phase mappings. When the user gazes at a specific flickering square, unique SSVEPs can be detected/decoded to infer user's gaze direction to the target character. However, modern BCI spellers are still constrained in regard of information transfer rate due to inaccurate decoding as well as high response latency.

\begin{figure}
	\includegraphics[width=\columnwidth]{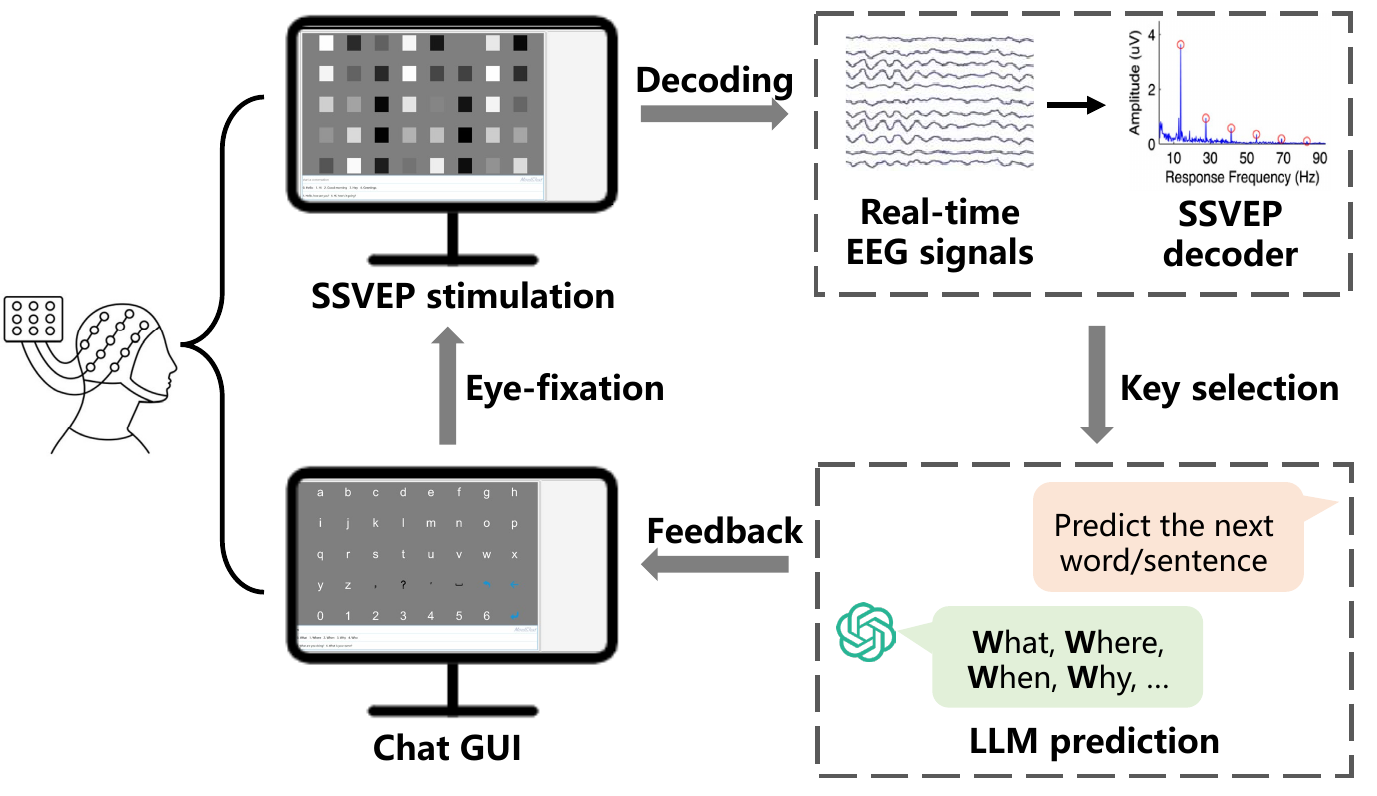}
	\caption{The framework of MindChat, a LLM-assisted SSVEP BCI speller.}
	\label{MindChat framework}
\end{figure}

A common solution to enhance BCI efficiency is to improve decoding performance from both accuracy and latency aspects. Although numerous efforts have been made to enhance decoding accuracy and latency in the past decade \citep{Liu2023,Nakanishi2018}, it still remains an obstacle to real-life applications in humans. On the one hand, it is noted that even a 95\% decoding accuracy struggles to accomplish a simple text entry in a minimum step. For example, users are supposed to type characters letter-by-letter to compose a complete sentence (h-e-l-l-o-\_-w-o-r-l-d). In that case, each time the user types a character, there will be a 5\% error rate and the errors grow proportionally, resulting in cumbersome and laborious burdens on error correction. On the other hand, there exists an accuracy-latency trade-off in BCI decoding \citep{Gong2022}. Such trade-off is of particular concern in realistic chatting scenarios, where one utterance usually contains more than 30 characters. For a typical 1.5-s-long decoding window in SSVEP spellers, typing a single sentence can be extremely time-consuming and thus catastrophic for users due to fatigue and high cognitive load caused by the prolonged focus required to accommodate decoding accuracy.

In response to these challenges, we seek to leverage assistive text input methods that can seamlessly integrate with existing BCI spellers, significantly enhance spelling efficiency, and improve the overall user experience in realistic scenarios. Traditional input methods typically rely on data structures like directed word graphs, Tries, and priority queues to store and complete words, while edit distance is commonly used for spelling correction. Though cost-effective, they fall short of context associations and semantic nuances. Recent advances in large language models (LLMs) provide new opportunities as AI-assisted input methods. Trained on vast corpora of text, LLMs have demonstrated exceptional zero-shot and few-shot learning \citep{Zhao2023,Dong2024}, in which LLMs can follow instructions described in natural languages (known as 'prompts') to perform versatile and highly-specialized NLP tasks such as question answering \citep{Brown2020}, word prediction \citep{Cai2024}, and spell-checking \citep{Patel2021}. As regards BCI spellers, LLMs hold great potential to significantly enhance spelling efficiency by offering context-aware word completion/prediction and correction. Such textual assistance could be particularly beneficial for BCI users who are unable to use usual keyboard interaction, facilitating real-world textual communication regardless of the BCI decoding dilemma.

In this study, we present MindChat, an LLM-assisted SSVEP BCI speller that renders predictive words and sentences to enhance spelling efficiency in real-world chatting scenarios.
The main contributions of this study are as follows:
\begin{itemize}
	\item We introduce MindChat, the first SSVEP BCI speller that incorporates LLMs to enhance spelling efficiency by providing contextually appropriate word and sentence predictions.	
	\item We leverage prompt engineering to develop a novel LLM-based spelling method that is context-aware and plug-and-play while free from troublesome model training.
	\item We demonstrate the efficiency and usability of our LLM-assisted BCI speller through online copy-spelling tasks across four chatting scenarios including single-turn, multi-turn daily and healthcare dialogues.
\end{itemize}

\section{Related Work}

\subsection{BCI Spellers}
Since \citet{Farwell1988} first applied BCI to spelling, numerous advancements have been made in diverse BCI paradigms like SSVEPs \citep{Chen2015}, P300 \citep{Hu2024}, and motor imagery \citep{WangJH2023}, with a focus on improving decoding performance and GUIs. For instance, \citet{Liu2023} introduced a high-speed SSVEP speller using user-specific decoding algorithms. \citet{Li2015} boosted visual stimuli with chromatic and familiar face elements to strengthen elicited brain patterns. Similarly, \citet{Obeidat2015} refined row/column intensities in P300 spellers for better target selection. Despite these advances, BCI spellers still face challenges in terms of spelling efficiency and user experience. We address these challenges by integrating LLMs as intelligent spelling assistants.

\subsection{LLM-Enhanced Text Input}
Prompt engineering \citep{Sahoo2024} has emerged as an essential technique for unlocking the full potential of LLMs. Grounded in zero-shot \citep{Wei2022} and few-shot prompting \citep{Brown2020}, LLMs are capable of undertaking novel tasks without requirements for extensive training. Using crafted prompts, LLMs have been recently applied in varieties of text prediction tasks including abbreviation expansion \citep{Cai2022}, form filling for websites \citep{Aveni2023}, keyword-to-sentence generation \citep{Chen2024}, etc. LLMs' ability to predict text with minimal context cues makes them suitable for BCI spelling by reducing manual input requirements.

\section{Methods}

\subsection{Framework}
The overall framework of MindChat is shown in Figure~\ref{MindChat framework}. As for the GUI of MindChat, a virtual keyboard with 40 keys is presented including 26 alphabet keys for letters a to z, three punctuation keys for comma, query, and apostrophe, three function keys for undo, delete, and enter, seven index keys from 0 to 6 corresponding to seven suggested candidates, and one space key. Below the keyboard panel, there are three text lines showing the current text input, five candidate words, and two candidate sentences, respectively. The right panel displays dialogue history messages. As shown in the figure, typing is administrated trial-by-trial in a circular manner. At the beginning of each trial, the user decides and gazes at the target key that he wants to type. Next, SSVEP stimulation is presented with flicking squares of different frequencies (from 8 Hz to 16 Hz in a step of 0.2 Hz) and phases (0, 0.5$\pi$, 1.5$\pi$), during which EEG signals are synchronously recorded. After a duration of 1.5 s, an SSVEP decoding model is employed to infer the key selected. Then, a prompted LLM is leveraged to \textit{"predict the next word and sentence"} based on the current text input and dialogue context. In the sequel, the candidate words and sentences are updated and the next trial begins. The procedure proceeds recurrently until the 'enter' key is triggered.

\subsection{Prompting with LLMs}
\begin{figure}
	\includegraphics[width=\columnwidth]{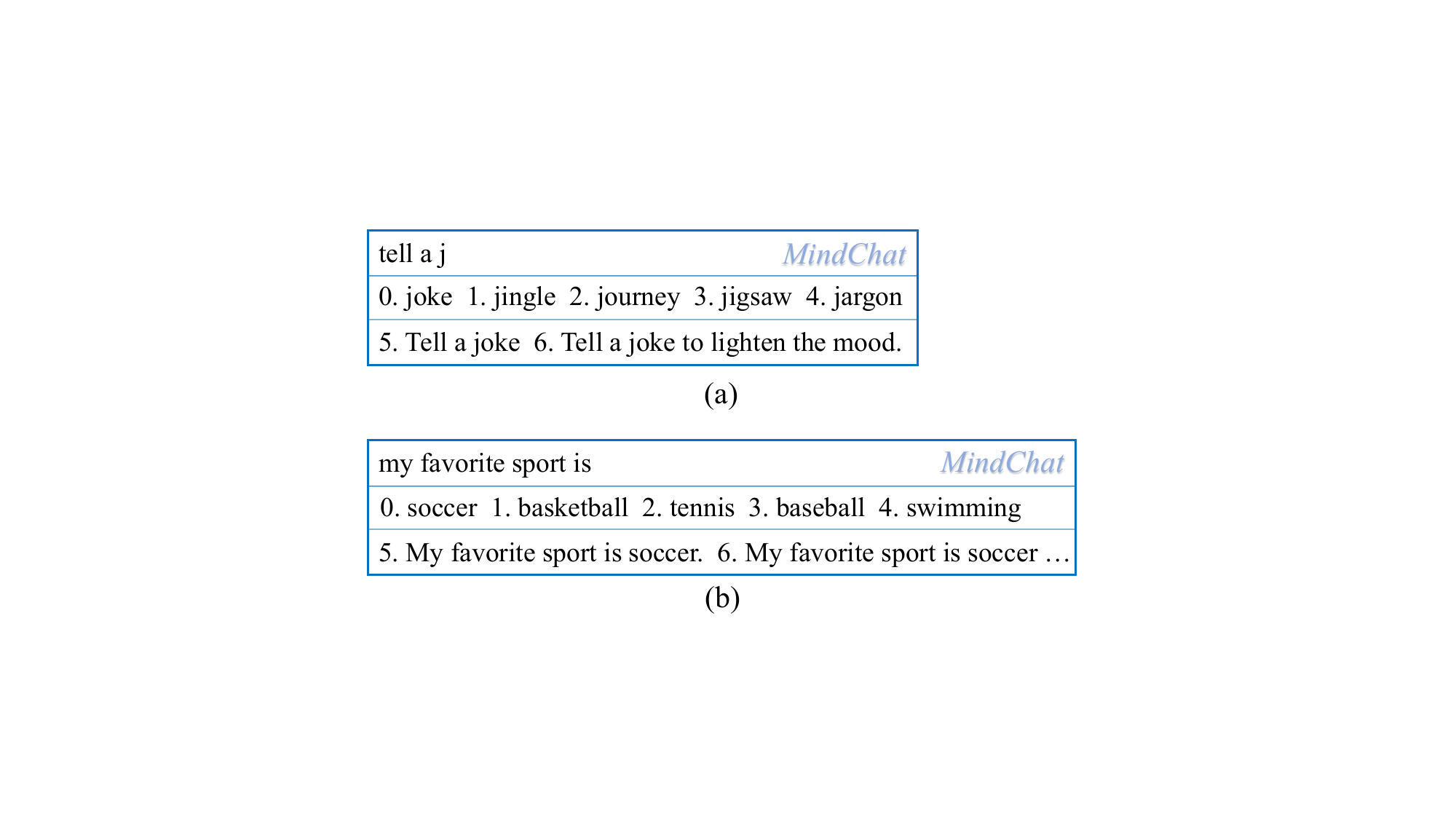}
	\caption{Example suggestions for word completion and word prediction, respectively.}
	\label{spelling suggestions}
\end{figure}

We employ OpenAI's GPT model with specific prompt engineering techniques to implement our LLM-assisted text input method. We primarily adopt GPT-4o (gpt-4o-2024-08-06) to generate predicted words and sentences via OpenAI's API \footnote{\url{https://api.openai.com/v1}}, complying with OpenAI's terms and policies. The prompt used for spelling suggestions can be found in Appendix~\ref{prompts}. Concretely, we prompt the LLM to perform either \emph{word completion} or \emph{word prediction} as well as \emph{sentence completion/prediction} based on the current text input and dialogue context. Two examples for both cases are shown in Figure~\ref{spelling suggestions}. In the upper panel, the utterance has been partially typed as 'tell a j'. Thus, 'j' is expected to be the prefix of the last word. As can be seen, the model properly completes the last word, providing reasonable word and sentence predictions. In the lower panel, the utterance is incomplete while the last word has been fully typed. As expected, the model predicts the next candidate words while contextually appropriate. To further ensure alignment with our instructions, we additionally leverage few-shot prompting to enhance LLMs' generation quality and robustness. The few-shot prompts are provided in Appendix~\ref{prompts}. Notably, we find few-shot prompting particularly benefits the word completion task as LLMs initially struggle to generate appropriate words according to the given prefix, especially models with weak NLP capability (e.g. GPT-3.5).

\begin{table}
	\centering
	\linespread{1}\selectfont
	\begin{tabular}{m{67pt}<{\centering} m{33pt}<{\centering} m{36pt}<{\centering} m{33pt}<{\centering}}
		\hline
		Dialogue Cat.	&	\# Utter.	&	\# Words	&	\# Char.	\\		
		\hline
		S-T daily	&	50	&	233	&	1534	\\
		S-T healthcare	&	50	&	214	&	1318	\\
		M-T daily	&	50	&	238	&	1298	\\
		M-T healthcare	&	50	&	230	&	1282	\\
		\hline
	\end{tabular}
	\caption{Statistics of the spelling dataset.}
	\label{data statistics}
\end{table}

\begin{table*}
	\centering
	\linespread{1}\selectfont
	\setlength{\tabcolsep}{5.2pt}
	\begin{tabular}{m{60pt}<{\raggedright}  cccccccc}
		\hline
		\multirow{2}{*}{Method} & \multicolumn{2}{c}{S-T daily} & \multicolumn{2}{c}{S-T healthcare} & \multicolumn{2}{c}{M-T daily} & \multicolumn{2}{c}{M-T healthcare} \\
		\cmidrule(r){2-3}\cmidrule(r){4-5}\cmidrule(r){6-7}\cmidrule(r){8-9}
		& time & keystrokes & time & keystrokes & time & keystrokes & time & keystrokes \\
		\hline
		Naive        & 124.68 & 30.78  & 106.5  & 26.36  & 104.24 & 25.80  & 102.94 & 25.48  \\
		DWG          & 103.92 & 17.62  & 96.50   & 16.40   & 90.64  & 15.44 & 96.12  & 16.32  \\
		GPT-4o       & 102.34 & \textbf{13.94}  & \textbf{86.14}  & \textbf{11.84}  & \textbf{68.74}  & \textbf{9.58}  & \textbf{62.12}  & \textbf{8.72}   \\
		GPT-4o-mini  & \textbf{101.16}	& 14.32  & 86.64  & 12.36  & 76.80   & 11.02 & 71.96  & 10.38  \\
		\hline
	\end{tabular}
	\caption{Performance of different assistive spelling methods across the four dialogue tasks. In offline analysis, the SSVEP decoding accuracy is set to 100\%.}
	\label{spelling performance}
\end{table*}

\subsection{Materials}
To validate the efficiency and effectiveness of our MindChat, a copy-spelling task is performed to simulate real-world chatting scenarios. Specifically, subjects are instructed to copy-spell the target utterance in the dialogue using MindChat. We craft our spelling dataset with four dialogue categories including single-turn, multi-turn daily and healthcare dialogues, broadly testing MindChat capability in diverse spelling conditions. The original dialogues used in this study are basically collected from two dialogue datasets, i.e. DailyDialog \citep{Li2017} and Alpaca \citep{Wang2023}. We further manually select one utterance in each dialogue as a spelling reference for the typing task, resulting in 50 target utterances for each dialogue category. As for multi-turn dialogues, dialogue histories are preserved as context information to facilitate context-aware predictions. Table~\ref{data statistics} shows the statistics of the constructed dataset. 
In the experiment, each subject types sixteen utterances with four utterances per dialogue category.

\section{Results}
\begin{figure}
	\includegraphics[width=\columnwidth]{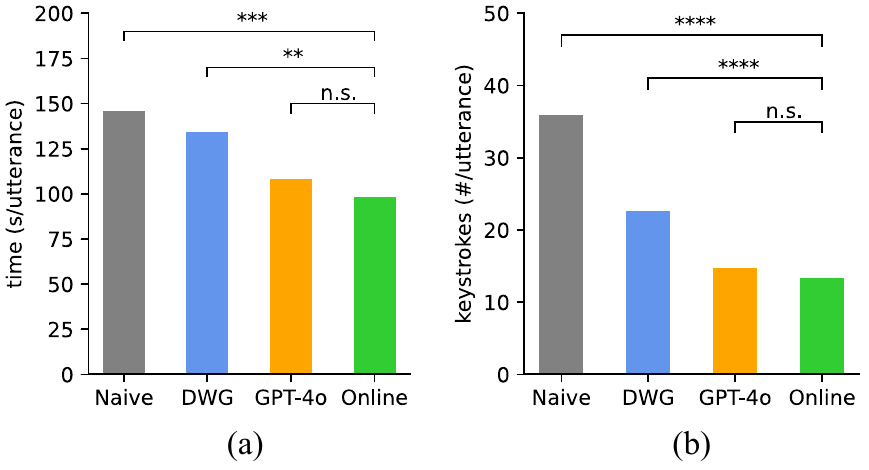}
	\caption{(a) Average spelling time per utterance. (b) Average number of keystrokes per utterance.}
	\label{online results}
\end{figure}

Figure~\ref{online results} shows the overall online performance compared with other pseudo-online evaluations. Specifically, a BCI speller with no assistive input method (naive mode), one using a traditional word completion method (directed word graph, DWG) \footnote{\url{https://github.com/seperman/fast-autocomplete}}, and our MindChat integrated with GPT-4o are additionally evaluated in a pseudo-online manner. To this end, we simulate the same spelling tasks appeared in online experiments while taking into account subjects' SSVEP decoding accuracy. During simulation, whenever an error occurs, the error needs to be corrected using the 'undo' command before spelling the remaining part of the utterance. The candidate words and sentences are prioritized to check if any of them fits the reference utterance. The performance is measured by averaged spelling time and keystrokes per utterance across subjects. As demonstrated, on average, MindChat achieves significantly better spelling efficiency, saving more than 60\% keystrokes (MindChat: 13.39, naive: 35.91, $P<0.001$) and 30\% spelling time (MindChat: 98.05, naive: 145.71, $P<0.0001$) compared with the naive BCI speller. Moreover, it also significantly outperforms a traditional text completion method which can be attributed to LLMs' powerful capability of context association. Interestingly, we observe a slight yet not significant improvement from pseudo-online to online situations when using the same GPT-4o assistance. This might be partially explained by subjects' adaptation to MindChat during online interaction, adopting more efficient shortcut strategies as well as improvements in decoding accuracy due to increased concentration. An example illustration of one kind of shortcut strategy is provided in Appendix~\ref{spelling example}.

We further conduct offline analysis of ideal spelling performance when the decoding accuracy reaches 100\%. As shown in Table~\ref{spelling performance}, MindChat integrated with GPT-4o consistently speedups spelling efficiency across four dialogue scenarios. Importantly, it exhibits more pronounced improvements in saved keystrokes for multi-turn dialogues compared with single-turn dialogues, with speedup rates of 64\% versus 55\%. This indicates MindChat effectively leverages context information to further enhance spelling predictions. We also evaluate a lightweight model GPT-4o-mini which is economical and fast-to-response. As seen, the prompted GPT-4o-mini is also capable of assisting BCI typing, achieving comparable results as GPT-4o. This indicates the transportability of our crafted LLM prompts for spelling predictions.

\section{Conclusion}
In this paper, we have presented a novel LLM-assisted BCI speller named MindChat. Up to our knowledge, MindChat is the first SSVEP-based BCI speller that incorporates LLMs to enhance spelling efficiency and overall user experience. 
For a group of ten BCI-naive subjects, online dialogue tasks demonstrate significant improvements compared with traditional BCI spellers.

\section*{Limitations}
The following limitations need to be further addressed. Firstly, in contrast to keystrokes which mainly depend on LLMs' predictions, spelling time is affected by multiple facets including SSVEP decoding window, LLMs' response time, and decision time. 
As for the decoding window, we presently set a 1.5-s-long window for the sake of robust and accurate decoding for BCI naive subjects, while a 0.5-s-long window is also demonstrated to be enough for well-trained BCI users. As for LLMs' response time, an average response time between 1 s to 2 s is observed in this study using the GPT-4o model. Future improvements could involve adopting lightweight LLMs or developing pruned LLMs optimized for spelling. As for decision time, an abundant time of 3 s is used for the consideration of given suggestions, we believe this could be effectively reduced with the proficiency of participants. Secondly, our method doesn't take into account users' specific linguistic profiles and personalities which contribute to customized and precise spelling suggestions. This could be achieved through personalized prompt engineering templates as well as user-tailored LLM fine-tuning. Thirdly, currently MindChat is only applied to English speakers. Future works can extend MindChat to a multilingual spelling assistant for broad populations.

\section*{Ethical Considerations}
This work involved human subjects or animals in its research. Approval of all ethical and experimental procedures and protocols was granted by the Ethics Committee of Zhejiang University.

\bibliographystyle{acl_natbib}
\bibliography{custom}

\clearpage
\appendix

\section{Participants}
We recruit a total number of 10 healthy adults (7 males and 3 females) aged between 20 and 26 years. All participants possess normal vision and are capable of communicating in English. All of them are BCI-naive subjects that have no prior experience with BCI spellers. Ethical approval for the study is obtained from the Ethics Committee of Zhejiang University. Written informed consent is obtained before the experiment.

\section{SSVEP Decoding}\label{ssvep decoding}
We follow the common practice in SSVEP decoding \citep{Mei2024} and a brief introduction of the decoding procedure is presented here. In our implementation, EEG data are acquired with a g.USBamp amplifier (g.tec, Inc.) at a sampling rate of 1200 Hz. Nine electrodes over parietal and occipital areas (Pz, PO1, PO2, POz, PO4, PO6, Oz, O1, and O2) are used to record SSVEPs, thus for each trial, EEG signals within the SSVEP stimulation period are extracted and represented by a 2D matrix in the shape of $C \times T$ wherein $C$ denotes EEG channels and $T$ denotes sampling points. EEG signals are first downsampled to 240 Hz and then band-pass filtered from 7 Hz to 70 Hz, eliminating irrelevant noise interference. We adopt filter-bank extended canonical correlation analysis (FBECCA) \citep{NAKANISHI2014} as the decoding algorithm which is representative as a variant of the classical CCA method. Its key principle is to measure the underlying correlation between two multi-dimensional variables $X$ and $Y$. In frequency-phase detection of SSVEPs, $X$ indicates multichannel EEG signals in multiple filter banks and $Y$ refers to reference template signals including sinusoidal templates and those derived from multiple SSVEP trials. By calculating the canonical correlation $\rho$ between multichannel SSVEPs and each reference signal. The reference template signal that maximizes the weighted correlation value is selected and thereby the corresponding target can be determined. The accuracies with different decoding methods are presented in Table~\ref{decoding accuracies}. It is observed that FBECCA yields an average decoding accuracy of 89\% comparable to those of FBDSP \citep{XiangLiao2007} and FBTRCA \citep{Nakanishi2018}. Nevertheless, the improvements brought by more advanced decoding algorithms are marginal. In recognition of the decoding bottleneck, we leverage an LLM-based spelling assistant and significantly enhance the efficiency and usability of current BCI spellers.

\section{Involved Prompts}\label{prompts}
\begin{figure*}
	\includegraphics[width=\linewidth]{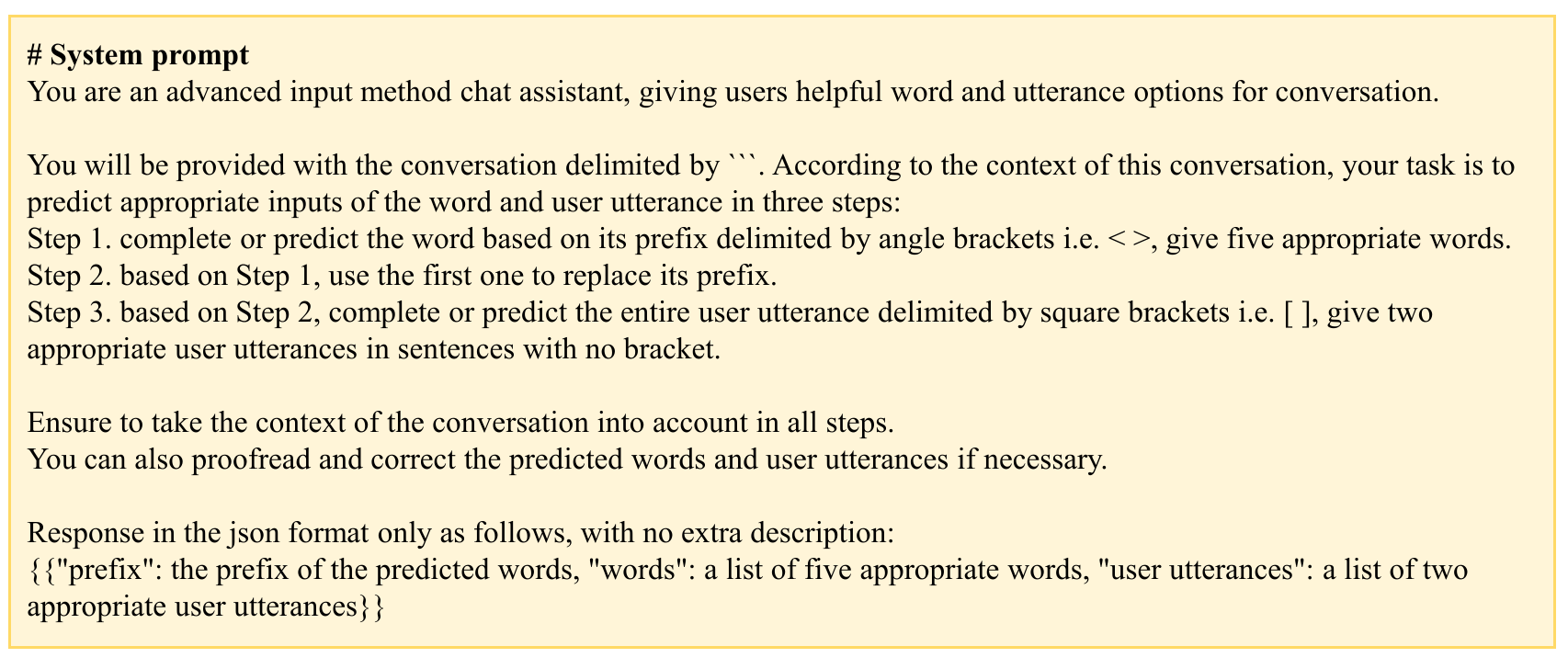}
	\caption{The detailed system prompt of word and sentence completions/predictions.}
	\label{system prompt}
\end{figure*}

\begin{figure*}
	\includegraphics[width=\linewidth]{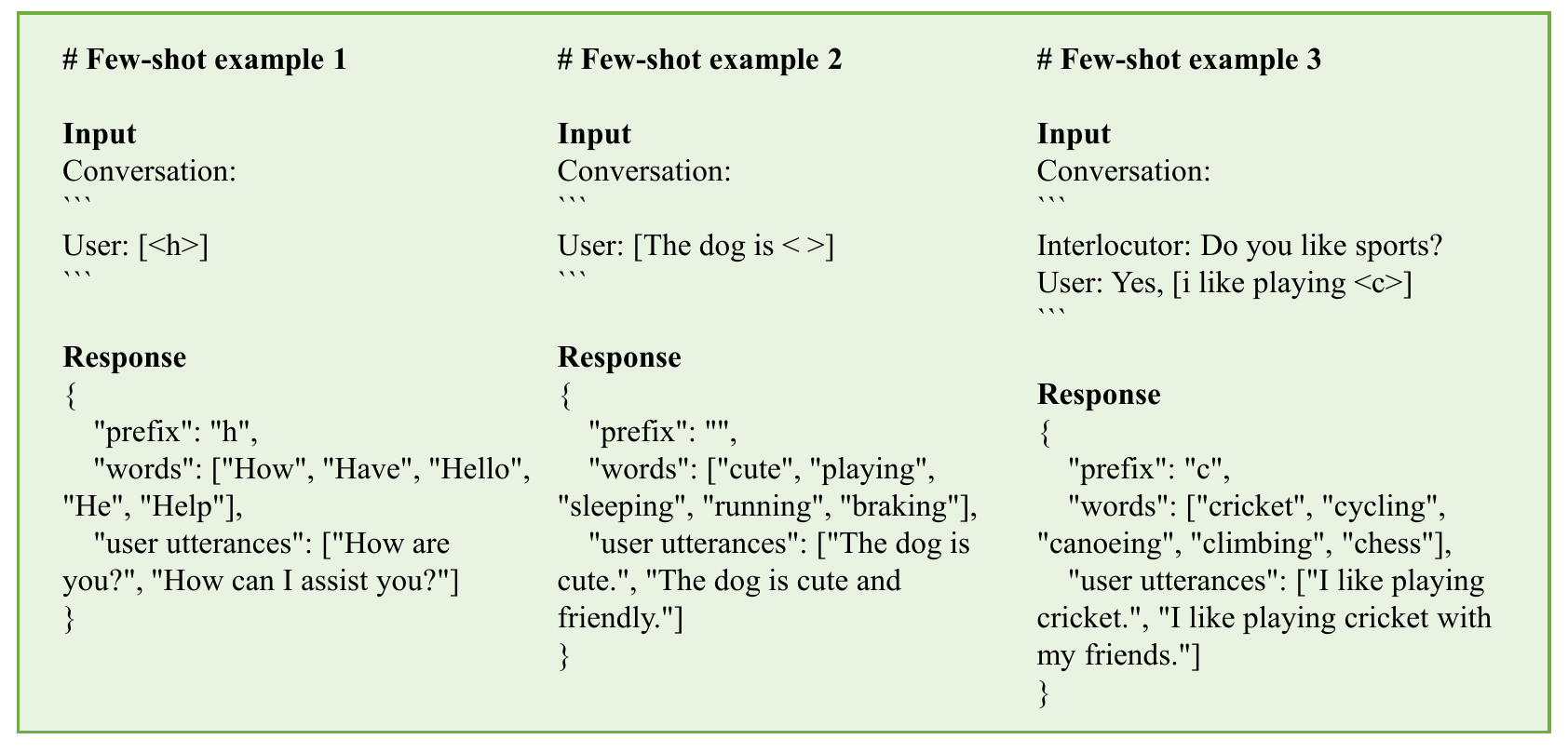}
	\caption{The few-shot prompting examples.}
	\label{few-shot prompts}
\end{figure*}

Figure~\ref{system prompt} is the essential prompt used for generating context-aware word and sentence predictions. Figure~\ref{few-shot prompts} illustrates the few-shot prompting examples we use to further guide LLMs' output toward system instructions.

\section{Spelling Example}\label{spelling example}
A detailed spelling example is presented in Table~\ref{spelling illustration}. We can observe that the user effectively utilizes suggested words and sentences to speed up the spelling of the reference utterance. Importantly, a unique shortcut strategy is employed by the user that further enhances spelling efficiency. Specifically, the user opts for first selecting a similar yet different sentence prediction (key index 39) and then deleting the last word (key index 32) so that the majority of the intended words can be preserved. Such shortcut strategy can be flexibly handled by users during practical spelling.

\begin{table}
	\centering
	\setlength{\tabcolsep}{3pt}
	\begin{tabular}{|l|c|}
		\hline
		\multicolumn{2}{|l|}{\textbf{Reference}: What's the best way to gain muscle?} \\
		\hline
		\textbf{Input text} & \textbf{Key index} \\
		\hline
		w & 23 \\
		what & \textcolor{blue}{33} \\
		What's & \textcolor{blue}{34} \\
		What's t & 20 \\
		What's the & \textcolor{blue}{33} \\
		What's the b & 2 \\
		What's the best way to proceed? & \textcolor{red}{39} \\
		What's the best way to & \textcolor{gray}{32} \\
		What's the best way to g & 7 \\
		What's the best way to gain & \textcolor{blue}{37} \\
		What's the best way to gain muscle? & \textcolor{red}{38} \\
		\hline
		\multicolumn{2}{p{220pt}}{\textbf{Key index}: (1-26) a-z; (27-30) comma, query, apostrophe, space; (31-32) undo, delete; (33-37) candidate words; (38-39) candidate sentences; (40) enter.}
	\end{tabular}
	\caption{A spelling example speeds up by word and sentence selections.}
	\label{spelling illustration}
\end{table}

\begin{table*}
	\centering
	\linespread{1}\selectfont
	\begin{tabular}{m{50pt}<{\centering} m{60pt}<{\centering} m{60pt}<{\centering} m{60pt}<{\centering} m{60pt}<{\centering}}
		\hline
		\makecell{Subject \\ No.}	&	FBSCCA	&	FBECCA	&	FBDSP	&	FBTRCA	\\		
		\hline
		s01	&	57.92	&	80.42	&	82.08	&	81.67 	\\
		s02	&	42.92	&	80.00	&	68.33	&	82.92	\\
		s03	&	55.83	&	96.67	&	99.58	&	99.17	\\
		s04	&	37.08	&	85.83	&	90.42	&	91.67	\\
		s05	&	56.25	&	87.92	&	92.08	&	89.58	\\
		s06	&	50.42	&	80.00	&	80.83	&	80.00	\\
		s07	&	79.17	&	99.58	&	100.0	&	100.0	\\
		s08	&	50.83	&	92.50	&	92.08	&	92.92	\\
		s09	&	42.92	&	91.67	&	97.50	&	99.17	\\
		s10	&	66.25	&	94.58	&	97.92	&	97.92	\\
		\hline
		Avg	&	53.96	&	88.92	&	90.08	&	91.50	\\
		\hline
	\end{tabular}
	\caption{SSVEP decoding accuracies (\%) using different methods.}
	\label{decoding accuracies}
\end{table*}

\end{document}